\documentclass[12pt]{iopart}
\usepackage{iopams}  

\usepackage{setstack}
\usepackage{epsf}
\usepackage{psfrag}
\usepackage[square,comma,numbers,sort&compress]{natbib}
\usepackage{graphicx}
\usepackage[squaren]{SIunits}

\def\dd{{\text{d}}}

\def\kesik{\protect\mbox{--\, --\, --}}
\begin{document}
\submitto{IEEE Trans Diel Elec Insul (\today)}

\title{On micro-structural effects in dielectric mixtures}
\author{Enis Tuncer}
\address{Applied Condensed-Matter Physics, Department of Physics, University of Potsdam, Am Neuen Palais 10, D-14469 Potsdam, Germany}
\ead{\tt enis.tuncer@physics.org}

\pagestyle{empty}
\begin{abstract}
The paper presents numerical simulations performed on dielectric properties of two-dimensional binary composites on eleven regular space filling tessellations. First,  significant contributions of different parameters, which play an important role in the electrical properties of the composite, are introduced both for designing and analyzing material mixtures. Later, influence of structural differences and intrinsic electrical properties of constituents on the composite's over all electrical properties are investigated. The structural differences are resolved by the spectral density representation approach. At low concentrations of inclusions (concentrations lower than the percolation threshold), the spectral density functions are delta-sequences, which corresponds to the predictions of the general Maxwell-Garnett mixture formula. At high concentrations of inclusions (close to the percolation threshold) systems exhibit non-Debye type dielectric dispersions, and the spectral density function differ from each other and that predicted by the Maxwell-Garnett expression. The analysis of the dielectric dispersions with an empirical formula also illustrate structural differences between the considered geometries, however, the information is not qualitative. The empirical formula can only be used to compare structures. The spectral representation on the other hand is a concrete method to characterize the structures of dielectric mixtures. Therefore, like in other spectroscopic techniques, a look-up table might be usefull to classify/characterize composite materials structure. This can be achieved by generating dielectric data for known structures as presented and emphasized in this study. Finally, the numerical technique, without any {\em a-priori} assumptions, for extracting the spectral density function is also presented.
\end{abstract}

\section{Introduction}

Composite materials are extensively used  for various mechanical, thermal and electrical application nowadays. Their electrical properties gain significant interest due to their possible applications as special materials, sensors and specially due to understanding the nature of already existing materials by utilizing electrical impedance measurements~\cite{MacDonald1987,Jonscher1983,Tuncer2002a}, such as in polymer physics and geophysics. Example of composites can be classified as man-made and natural mixtures~\cite{MiltonBook}. In the former class are composites (fiber, particulated and percolating), porous materials (gels, cellular solids and foams), colloids, microemulsions, block copolymers, concrete and even biological samples, which are some sort of emulsion in a continuous background of solution. Some of the natural mixtures are polycrystals, soil, sandstone, granular media, and biological materials such as wood, bone, blood, lungs, tissues, tumors {\em etc}. 

Although bulk structure of materials are not tailored yet, by the introduction of micro-electronics and nanotechnology ~\cite{Drexler,Feynman}, we are now able to manufacture desired structures using the lithography techniques. It is not yet possible to produce materials with three-dimensional structures, however, besides the micro-electronic mechanical systems (known as MEMS) attempts have been made to produce materials as sensors with lasers and photo-curable solutions~\cite{Torquato1998,Jacobs}. The process to miniaturize man-made equipment pushes scientist and engineers to come up with new manufacturing ideas. Currently most of the research has focused on the surface manipulation of materials, however, the electro-kinetic~\cite{Morgan0022-3727-36-20-023,Morgan0022-3727-33-6-308,Morgan0022-3727-30-11-001,Hughes0957-4484-11-2-314,Hughes0022-3727-29-2-029} and acoustic techniques would infact create the opportunity to tailor the bulk properties and structures of material composites. If this is possible, then computer simulations~\cite{BBReview,Brosseau1,BrosseauQue,Boudida2,Brosseau2,Boudida,BrosseauBoulic,Sareni1,sar97,sar97mag,TuncerPhysD,TuncerAcc1,Tuncer2002a,Tuncer2002b,Tuncer-CEIDP01,Tuncer2001a,Tuncer2001elips,Tuncer2002elec,TuncerPhD,TuncerRandom,Tuncer1998b,KarkkainenPhD,Karkkainen,Pekonen1999,Pelster2001} and dielectric mixture formulas~\cite{LL,SihvolaBook,Landauer1978,Pier6,Tuncer2001a,Karkkainen,Merrill,Sihvola2000,Friedman1998} would be used to predict the behavior of materials before and to characterize after the manufacturing.

The analytical or empirical formulas are general expressions for the mixtures, they are only useful, would yield valuable information, when the composite lacks complexity in its topology and one of the phases has low volume fractions\cite{TuncerPhysD,TuncerAcc1,Tuncer2002b,TuncerLic,TuncerRandom}. In the last couple of decades with the improvements in the numerical codes and computation speed, numerical solutions to obtain the electrical properties have been widely used~\cite{BBReview,Brosseau1,BrosseauQue,Boudida2,Brosseau2,Boudida,BrosseauBoulic,Sareni1,sar97,sar97mag,TuncerPhysD,TuncerAcc1,Tuncer2002a,Tuncer2002b,Tuncer-CEIDP01,Tuncer2001a,Tuncer2001elips,Tuncer2002elec,TuncerPhD,TuncerRandom,Tuncer1998b,KarkkainenPhD,Karkkainen,Pekonen1999,Pelster2001}. The numerical calculations overcome the disadvantages of the previously mentioned items, however, they have also handicaps such as the size of the problem, the ratio of the largest to smallest unit in the structure, and numerical errors. 

One of the novel features in composites, unfortunately not used widely, was proposed by Bergman~\cite{Bergman1,Bergman1978,Bergman4}, which is called {\em the spectral representation theory}. In this theory if the electrical properties of the constituents of a binary mixture are known, then the contribution of the topology, or in other words the geometrical arrangement of constituents, can be extracted from the information hidden in the Maxwell-Wagner-Sillars polarization~\cite{Maxwellbook,Wagner1914,Sillars1937,Tuncer2002a}. In addition one can used this approach to resolve the dielectric permittivity of a unknown material in a host medium, if a test has been performed with a known material with the same geometrical shape and distribution as the unknown one in the host medium. This can be achieved by using the same spectral density function for the unknown material and reverse engineer to obtain the dielectric properties of the unknown material. Previously, researcher have proposed cumbersome analytical expressions, which are based on the adopted spectral density functions~\cite{Stroud,GhoshFuchs}. Here, it is  not only discussed how to extract the unknown spectral density function but it is applied  to extract structural differences of two-dimensional structures, which are unidirectional fibers in a host medium in three-dimensions. The extraction is based on a numerical technique developed by the present author~\cite{Tuncer2004a,Tuncer2000b}. It is based on the Monte Carlo integration hypothesis and the constrained-least-squares algorithm. There are no  {\em a-priori} assumptions in the procedure as the previous researchers proposed. The method is tested on eleven two-dimensional space-filling tessellation lattices/structures~\cite{StructureBook}. The dielectric permittivity of structures are calculated by the finite element method~\cite{TuncerPhysD,TuncerAcc1,Tuncer2002a,Tuncer2002b,Tuncer2001a,TuncerRandom,Tuncer-CEIDP01}. The influence of the intrinsic electrical properties of the constituents on the dielectric relaxations are also investigated.

\section{Background}
\subsection{Factors influencing composite properties}

Composites are new sort of materials in which we have the possibility to tailor their properties for desired applications. The tailoring can be as in the form of ({\em i}) combining the intrinsic properties of the constituents, ({\em ii}) creating a new non-existing property, such as, interfacial polarization in mixtures, electro-mechanical activity in porous materials, meta-materials (negative permittivity/permeability and Poisson's ratio materials), and ({\em iii}) nullifying a property, this is achieved by destroy the structure of the continuous matrix. 

If we focus on the significant items influencing the dielectric properties of composites, they can be briefly listed as follows;
\begin{itemize}
\item intrinsic properties of constituents, {\em i.e.} ohmic conductivity $\sigma$, high frequency dielectric permittivity $\epsilon$ and dielectric susceptibility $\chi(\omega)$. The first two quantities are material constants, the latter quantity contains the dynamic properties of the constituents. The complex dielectric permittivity of a material ${\overline \varepsilon}$ can in general be expressed as a function of the angular frequency $\omega$
  \begin{eqnarray}
    \label{eq:1}
    {\overline \varepsilon}(\omega)=\epsilon+{\overline \chi}(\omega)+\sigma(\imath\varepsilon_o\omega)^{-1}
  \end{eqnarray}
where the over-lines represent the quantities are complex numbers. The permittivity is a function of angular frequency $\omega$, and losses due to ohmic condiutivity $\sigma$ contribute to the imaginary part of the permittivity. $\varepsilon_0$ is the permittivity of the free space, $\varepsilon_0=8.854\ \farad\reciprocal\meter$ and $\imath=\sqrt{-1}$. Keep in mind that the permittivity of materials are quantized meaning that the permittivity axis is not continuous. This is important while designing high and low permittivity materials due to selection of constituents.  The conductivity of materials on the other hand can be altered by doping. 
\item concentrations or volume fractions of constituents. This case is widely used in the industry to make semi-conducting/conducting polymers by using carbon-black powder, which is a dopant. 
\item size, shape, orientation and distribution of inclusions (if one of the phases is in particulated form). Influence of particle size on electrical properties have been investigate in percolating systems. Although it is advantageous to use mono-dispersed spherical inclusions to design composites, the mechanical compatibility is reduced due to adhesion. Therefore, usually fillers with arbitrary shapes and largest surface area are selected. Still, there is no concrete examples of bulk materials, in which orientation and distribution of the inclusions are controlled.
\item mixture topology. There is no concrete examples of controling the geometrical arrangements of constituents other than micro-electronics designs for lateral structures and arrangement of nano-scale particles on surfaces. The current study is focused on this aspect of mixtures, by performing numerical simulations on ``ideal composites'', in which their electrical and structural properties are manupulated. When the inclusion phase is expensive control of the distribution of them can in principle reduce materials cost or toxic vast without affecting the over-all composite properties.
\item inter-phase/interface between the phases. The interface between phases is an interesting topic since it is not clear how the transition from one phase to the other takes place. One of the open questions is the presence of an inter-phase and its dimensions and its influence on the overall composite properties. For example, samples prepared with surface modified fillers exhibit changes in the dielectric dispersions~\cite{Tuncer-CEIDP00}. Moreover, as the particles or phases agglomerate and the sizes of the phases decrease the surface to volume ratio of inclusion phase increase and the system is dominated with the surface effect.
\end{itemize}

\subsection{Importance of the mixture topology}\label{sec:import-mixt-topol}
\begin{figure}[t]
  \centering
  \includegraphics[width=4in]{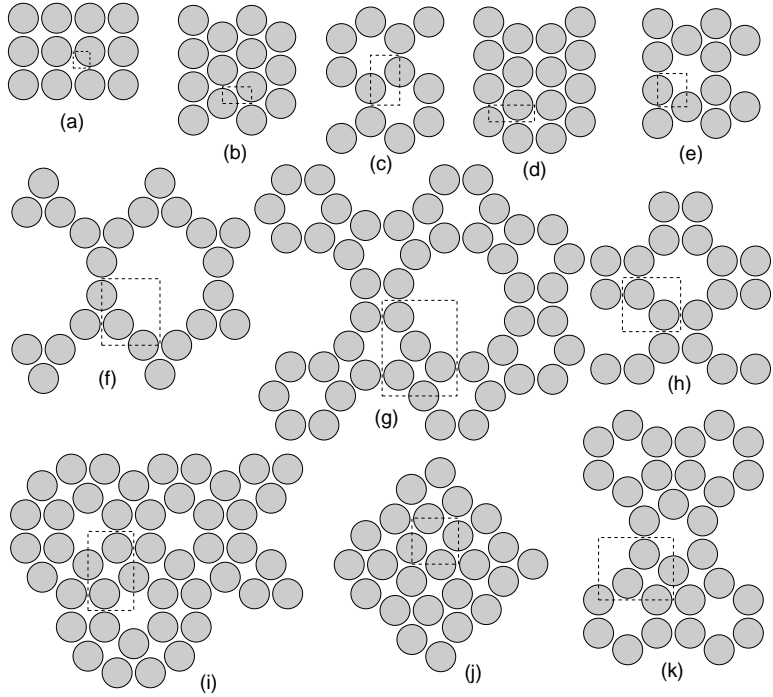}
\caption{Two-dimensional structures considered in the simulations. The first three structures (a), (b) and (c) are the well-known square, triangular and hexagonal(honeycomb) networks, respectively. The others are generated from various space-filling tessellations. The regions marked with rectangles are the computation domains used in the finite element analysis.}  \label{fig:geometries}
\end{figure}

As an example, if we are asked to come up with a novel materials design for a specific application, we must ponder about the items listed in the previous section. Neglecting the effect of the inter-phase/interface, once we have decided the components, matrix and inclusion phases, inclusion concentration and their size, we can obtain different electrical properties for the mixture just by considering different geometries--distribution of inclusions\cite{Tuncer1998b}. For this purpose, eleven different space-filling two-dimensional geometries are considered as presented in Fig.~\ref{fig:geometries}. The filler disks in two-dimensions are actually infinite unidirectional fiber inclusions in three-dimensions with the cross-sections in Fig.~\ref{fig:geometries}. The structural information on the geometries are presented in Table~\ref{tab:table1}. It is worth mentioning that some of the structures have similar limiting volume fractions $q_{\sf m}$, however, the coordination number $Z$~\cite{Zallen} (number of closest neighbors) are different. The limiting concentrations  are actually the percolation threshold~\cite{Percolation,Clerc1990,Kirkpatrick3,Kirkpatrick1} for a given regular geometry. Close to the percolation threshold local field is deformed from that of the average one (predicted by the effective medium approximation). The individual inclusions polarization is transformed from dipole approximation to  multipoles. At the some concentration level the effective medium theories become unapplicable.

If we return back to the two-dimensional structures by taking into consideration Fig.~\ref{fig:geometries} and Table~\ref{tab:table1}. The geometries in Fig.~\ref{fig:geometries}g and \ref{fig:geometries}h have similar limiting concentrations ($q_{\sf m}\sim0.55$) and coordination numbers. In such a case one can for example investigate the influence of the second nearest neighbors on the dielectric permittivity. Figs.~\ref{fig:geometries}a and \ref{fig:geometries}d have similar limiting concentration ($q_{\sf m}\sim0.78$) but coordination numbers. The geometries in Fig.\ref{fig:geometries}c,  \ref{fig:geometries}i and  \ref{fig:geometries}k  have on the other hand besides having similar limiting volume fractions ($q_{\sf m}=0.61$), their coordination numbers differ from each other, introducing an interesting case to study the influence of structural differences and number of the closest neighbors on electrical properties.  These significant differences are emphasized in the text by analyzing the electrical properties and the resulting spectral density functions of these structures with varied intrinsic properties of the constituents.

\begin{table}[t]
  \caption{Structural information regarding the considered two-dimensional geometries in Fig.~\ref{fig:geometries}; the coordination number $Z$, limiting volume fraction $q_{\sf m}$, the size of the computational geometry $a\times b$ in the maximum filler particle radius $r_{\sf m}$  and the fraction of the filler in the computation domain $n_{\sf e}$.} \label{tab:table1}
  \lineup
  \begin{indented}
  \item[]  \begin{tabular}{crrr@{$\times$}lr}
      \br
      Fig.~1 & $Z$ & $q_{\sf m}$ & $a$&$b~[r_{\sf m}]$ & $n_{\sf e}$ \cr 
      \mr 
      a & $4$ & $0.78$ & $1$&$1$ & $\scriptstyle{\frac{1}{4}}$ \cr
      b & $6$ & $0.90$ & $\sqrt{3}$&$1$ & $\scriptstyle{\frac{1}{2}}$ \cr
      c & $3$ & $0.61$ & $ \sqrt{3}$&$3$ & $1$ \cr
      d & $5$ & $0.77$ & $1+\sqrt{3}$&$1$ & $\scriptstyle{\frac{3}{4}}$ \cr
      e & $4$ & $0.68$ & $\sqrt{3}$&$2$ & $\scriptstyle{\frac{3}{4}}$ \cr
      f & $3$ & $0.45$ & $2+\sqrt{3}$&$2+\sqrt{3}$ & $2$ \cr
      g & $3$ & $0.55$ & $3+\sqrt{3}$&$2+2\sqrt{3}$ & $4\scriptstyle{\frac{1}{2}}$ \cr
      h & $3$ & $0.54$ & $2+\sqrt{2}$&$2+\sqrt{2}$ & $2$ \cr
      i & $5$ & $0.61$ & $1+\sqrt{3}$&$3+\sqrt{3}$ & $3$ \cr
      j & $5$ & $0.84$ & $1+\sqrt{3}$&$1+\sqrt{3}$ & $2$ \cr
      k & $4$ & $0.61$ & $1+2\sqrt{3}$&$2+\sqrt{3}$ & $3\scriptstyle{\frac{1}{4}}$ \cr
      \br
    \end{tabular}
  \end{indented}
\end{table}
\section{Determining topology contributions}
\subsection{Spectral (geometric) density function}
Bergman~\cite{Bergman1,Bergman1978,Bergman4} has proposed a mathematical way for representing the effective dielectric permittivity $\varepsilon_{\sf e}$ of a binary mixture as a function of permittivities of its constituents, $\varepsilon_1$ and $\varepsilon_2$, and an integral equation, which includes the geometrical contributions. It is called {\em the spectral density representation}. The representation utilizes the Maxwell-Wagner-Sillars polarization and the geometrical contributions, which are calculated as a function of depolarization factors~\cite{Tuncer2001d,Sillars1937,Steeman1992,Steeman90,Vila1992}.

After the introduction of non-destructive measurement techniques and systems, such as electrical~\cite{Jonscher1983,McCrum} or acoustic impedance spectroscopy~\cite{McCrum}, the impedance of materials (either pure or composite) could be recorded for various frequencies $\nu$. Then, the frequency could be used as a probe to obtain microstructural information with the application of the spectral density representation~\cite{Stroud,GhoshFuchs,Stroud1999,Day,Lei2001,Gonc2003,Cherkaev2003}. This can only be achieved if ({\em i}) no influence of $\nu$ on the geometrical arrangement of phases is present [the geometry should be static at each frequency, meaning that no piezoelectricity exists in the constituents, and the elastic properties of the phases should be the same or similar to each other otherwise there would be a nonzero displacement vectors (deformation) in the composite], ({\em ii}) the intrinsic properties of phases are known as a function of $\nu$. Numerical~\cite{Day,Gonc2003,Cherkaev2003} and analytical~\cite{Stroud,GhoshFuchs,Stroud1999} approaches have been used and proposed to resolve the spectral density function for composites. Even a relation between the spectral density function and  distribution of relaxation times have been ascribed~\cite{Lysne1983,Roussy1992}. The distribution of relaxation times approach\cite{Tuncer2004a,Tuncer2000b} has previously been applied  to illustrate the differences in random and regular structures~\cite{Tuncer2002b,Tuncer2002elec}. Although numerical approaches could be prefered over the analytical ones, which are empirical expressions and are not universal, they solve a nontrivial---ill-posed---inverse problem~\cite{Cherkaev2003}. Here, the same numerical method  as the distribution of relaxation times is applied to extract the the spectral density function of a binary mixture. The method is based on the Monte Carlo integration and constrained-least-squares algorithms.

\begin{figure}[t]
  \centering
  \includegraphics[]{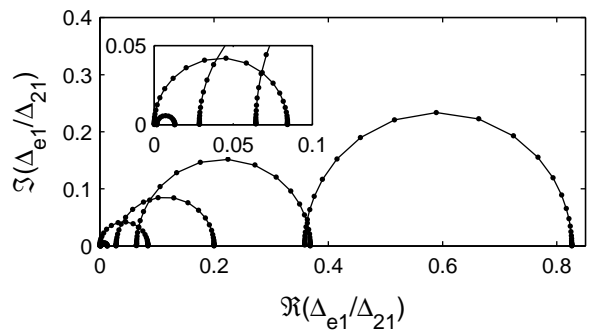}
  \caption{Parametric plot of the scaled mixture permittivity. The symbols are the analytical model of Maxwell-Garnett equation, and the solid lines (\full) are the values calculated from the spectral functions obtained from the proposed numerical method.  The semi-circles from large to small corresponds to $q_2=\{0.95,\, 0.7,\,0.5,\, 0.3,\,0.05\}$, respectively. The inset is the enlargement of the values close to the origin for $q_2=\{0.05,\,0.30\}$.}
  \label{fig:normalized}
\end{figure}

For a binary composite system with constituent permittivities $\varepsilon_1$ and $\varepsilon_2$, and concentrations $q_1$ and $q_2$, ($q_1+q_2=1$), and with an effective permittivity $\varepsilon_{\sf e}$, the spectral density representation is expressed as,
\begin{eqnarray}
  \label{eq:delta}
  \Delta_{{\sf e}i}/\Delta_{ji} - A_j = \int_0^1 {\sf g}_j(x)\,[1+\varepsilon_i^{-1}\Delta_{ji}x]^{-1}\, \dd x
\end{eqnarray}
where, $\Delta_{ij}=\varepsilon_i-\varepsilon_j$, and is complex and frequency dependent. $A_j$ is a constant, and depends on the concentration and structure of the composite. Several different notations have been used in the literature see Refs.~\cite{Stroud},\cite{GhoshFuchs} and \cite{Day}. Here, we rearrange the expression by Ref.\cite{GhoshFuchs}, and obtain a similar one those used by Refs.~\cite{Stroud} and ~\cite{Day}. The spectral density function is ${\sf g}(x)$, and it is sought by the presented procedure. The spectral density function satisfies $\int{\sf g}_j(x)\dd x=q_j$~\cite{Bergman1978,Gonc2003} and $\int x {\sf g}_j(x)\dd x=q_j \, q_i/d$, where $d$ is the dimension of the system. The shape of the inclusions in a matrix can also be related to $d$~\cite{Sillars1937,Tuncer2001d}. Finally,  $x$ is called the depolarization factor.

The numerical procedure is briefly as follows: first the integral in Eq.~(\ref{eq:delta}) is written in a summation form over some number of randomly selected (known) $x_n$-values, $x_n\in [0,1]$. This  converts the non-linear problem in hand to a linear one with only ${\sf g}_{j_{n}}$ values  being unknowns. Later, a constrained-least-squares is applied to get the corresponding ${\sf g}_{j_{n}}$-values:
\begin{eqnarray}
  \label{eq:minn}
  \min||{\bf \Delta}-{\bf K}{\sf g}_{j_{n}} ||_2 \quad {\rm and} \quad {\sf g}_{j_{n}}\ge 0
\end{eqnarray}
where ${\bf \Delta}$ is the left-hand-side of Eq.~(\ref{eq:delta}), and ${\bf K}$ is the kernel-matrix, $[1+\varepsilon_i^{-1}\Delta_{ji} x_n]^{-1}$. When this minimization is run over-and-over with new sets of $x_n$-values, most probable ${\sf g}_{j_{n}}$-values are obtained. For a large number of minimization loop, actually the $x$-axis becomes continues---the Monte Carlo integration hypothesis. Finally, the weighted distribution of ${\sf g}_{j_{n}}$ versus $x_n$ leads ${\sf g}(x)$. This is achieved by dividing the $x$-axis in channels and averaging ${\sf g}_{j_{n}}$ in each channel.

Application of the numerical procedure to the Maxwell-Garnett expression~\cite{Maxwell_Garnett} should yield delta function distributions for ${\sf g}(x)$~\cite{Stroud,Gonc2003}. The dielectric function for a $d$-dimensional (or composite with arbitrary shaped inclusions) Maxwell-Garnett composite is defined as 
\begin{eqnarray}
  \label{eq:4}
  \varepsilon_{\sf e}=\varepsilon_1 [1+ d\, q_2\, \Delta_{21}\,
  (q_1\,\Delta_{21} + d\, \varepsilon_1)^{-1}].
\end{eqnarray}
The resulting the spectral density function is then,
\begin{eqnarray}
  \label{eq:3}
  {\sf g}_j (x)=\delta[x-(1-q_j)/d].
\end{eqnarray}
We choose the following values for dielectric functions of the phases: $\varepsilon_1=1-\imath\,(100\varepsilon_0\omega)^{-1}$ and $\varepsilon_2=10-\imath\,(\varepsilon_0\omega)^{-1}$. The left-hand-side of Eq.~(\ref{eq:delta}) without the constant $A_2$ is plotted for a 3-dimensional composite ($d=3$ corresponding  spherical inclusions) in Fig.~\ref{fig:normalized} as a parametric plot of the imaginary part of $\Delta_{{\sf e}1}/\Delta_{21}$ against its real part. The graph is a semi-circle for the Maxwell-Garnett expression. In the figure, five different concentration levels are plotted, $q_2=\{0.05,\,0.3,\,0.5,\,0.7,\,0.95\}$, the inset shows the enlargement close to the origin, which illustrates the low concentrations, $q_2=\{0.05,\,0.3\}$. The size of the semi-circles are proportional to the concentration of Phase 2. The analyses performed on the scaled effective permittivity, Eq.~(\ref{eq:delta}), with the help of the applied method yield the solid lines (\full) in the figure.

\begin{figure}[t]
  \centering
  \includegraphics[]{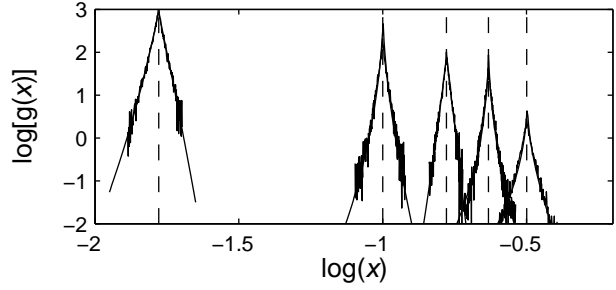}
  \caption{Calculated spectral density distributions, which correspond to delta sequences. The spectral functions from left to right corresponds to $q_2=\{0.95,\, 0.7,\, 0.5,\, 0.3,\,0.05\}$, respectively. The corresponding (calculated) $A_2$ values are $\{0.002,\,0.012,\,0.029,\,0.064,\,0.358\}$, respectively, for the considered concentrations. The dashed lines (\kesik) show the positions of the actual delta-functions for the Maxwell-Garnett expression.}
  \label{fig:gxMG}
\end{figure}

The corresponding ${\sf g}(x)$ are plotted in Fig.~\ref{fig:gxMG} on a log-log scale. In the figure, the expected locations of ${\sf g}(x)$ from Eq.~(\ref{eq:3}) are also shown with dashed lines (\kesik). The ${\sf g}(x)$-distributions obtained are analyzed by the L{\'e}vy distribution~\cite{Levy}, which generates a delta-sequence~\cite{Butkov}. The solid lines (\full) illustrate the appropriate L{\'e}vy distributions. Various parameters from the statistical analyses of the spectral density function, and their expected values are presented in Table~\ref{tabtableMG}. The concentration values, $\overline{q_2}$, calculated from the integration of ${\sf g}(x)$ without {\em a-priori} assumption are $<1\%$ for the considered high concentrations, and it is around 5\% for the lowest concentration, $q_2=0.05$. The localization parameter for the depolarization factor $x$, which is the most probable depolarization value, can be calculated by the integration of $[1-{\sf g}(x)]/d$ or with the help of statistical analysis. The estimated depolarization factors $\overline{x}$ are within $<1\%$ of the actual values stated by the proposed analytical expression~\cite{Stroud,Gonc2003}. Finally, the product of the concentrations $\overline{q_1 q_2}$, the integration of $3x{\sf g}(x)$, calculated have also very good agreement with those values expected from the definitions of the spectral density representation.

\begin{table}[t]
\caption{Comparison between the results of the proposed numerical approach and those of the  L{\'e}vy statistics and the given analytical the spectral density function for the Maxwell-Garnett effective permittivity expressions for various concentrations. The bars on the quantities indicate that they are calculated from the numerical results.}\label{tabtableMG}

\begin{indented}
\lineup
\item[]\begin{tabular}{@{}*{8}{r}}
\br
$q_2$   & 
$\overline{q_2}$ $^{\rm a}$ &  
$\overline{x}$ $^{\rm b}$ & 
$q_1/d$ $^{\rm c}$ & 
$A_2{_{in}}$ $^{\rm d}$ &
$A_2{_{out}}$ $^{\rm e}$ &
$\overline{q_1 q_2}$ $^{\rm f}$ &
$q_1 q_2$ $^{\rm c}$ \cr 
\mr
0.05  & 0.053 & 0.318 & 0.316 & 0.002 & 0.002 & 0.057 & 0.048\cr
0.30  & 0.301 & 0.234 & 0.233 & 0.012 & 0.013 & 0.213 & 0.210\cr
0.50  & 0.050 & 0.167 & 0.167 & 0.029 & 0.029 & 0.249 & 0.249\cr
0.70  & 0.704 & 0.100 & 0.100 & 0.064 & 0.064 & 0.213 & 0.280\cr
0.95  & 0.951 & 0.017 & 0.017 & 0.358 & 0.359 & 0.051 & 0.048\cr
\br
\end{tabular}
\item[] $^{\rm a}$ Calculated using the resulting ${\sf g}_2(x)$. Known from the definition of ${\sf g}_j(x)$---integral $\int_0^1{\sf g}_j(x)\dd x$ is equal to this value.
\item[] $^{\rm b}$ The localization parameter for the calculated L{\'e}vy distribution. The shape parameters and the amplitude of the L{\'e}vy distributions are disregarded. 
\item[] $^{\rm c}$ Known from the definition of the spectral density function for the Maxwell-Garnett expression, Eq.~(\ref{eq:3}).
  \item[] $^{\rm d}$ $A_2$-value calculated before the numerical procedure using Eq.~(\ref{eq:delta}).
\item[] $^{\rm e}$ Mean $A_2$-value calculated during each Monte Carlo integration step in  the numerical procedure, Eq.~(\ref{eq:minn}).
\item[] $^{\rm f}$ Calculated using the resulting ${\sf g}_2(x)$ and $x$-values. Known from the definition of ${\sf g}_j(x)$---the values is equal to the integral $\int_0^1 3x{\sf g}_j(x)\dd x$.
\end{indented}
\end{table}




\subsection{Numerical calculations}
\begin{figure}[t]
\centering
  \begin{tabular}{lr}
  \includegraphics[width=2.5in]{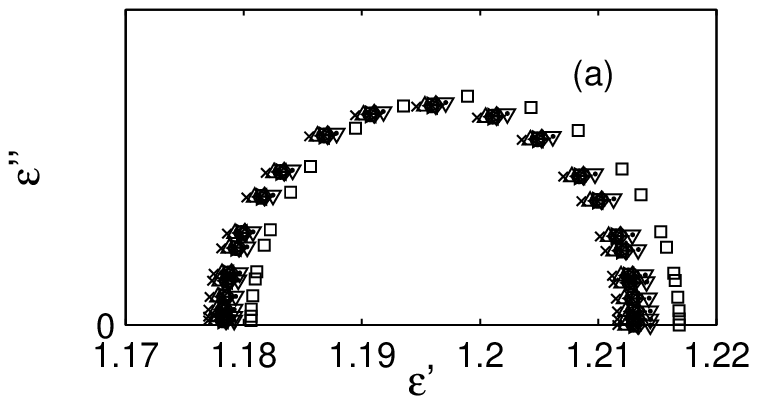}&
  \includegraphics[width=2.5in]{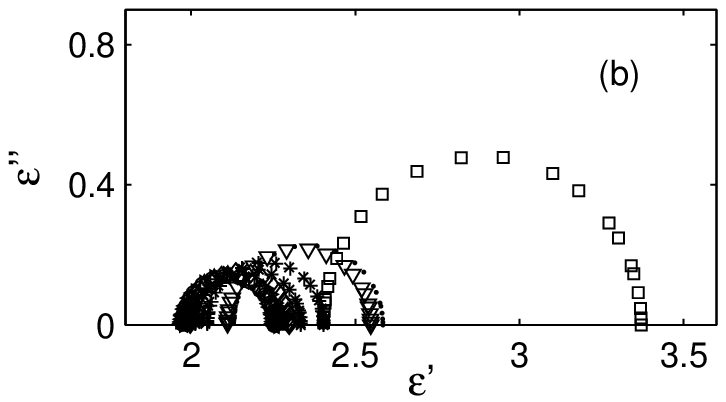}\\
  \includegraphics[width=2.5in]{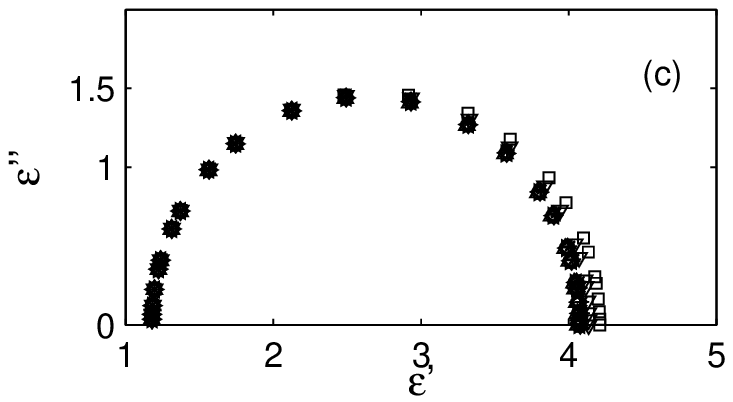}&
  \includegraphics[width=2.5in]{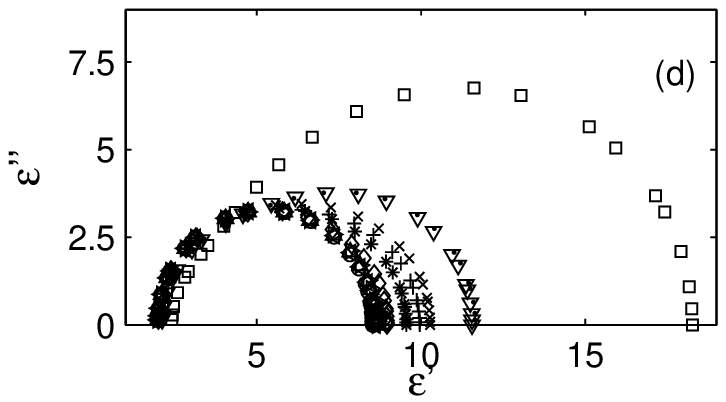}
  \end{tabular}
\caption{Parametric (Cole-Cole) plots  of dielectric permittivities with varied material parameters at different concentrations; match-composite at (a) $q=0.1$ and (b) $q=0.4$, and reciprocal-composite (c) $q=0.1$ and (d) $q=0.4$. The symbols represent the structures in Fig.~\ref{fig:geometries} as follows; (a:$\rhd$, (b:$\circ$), (c:$+$), (d:$\lhd$), (e:$\star$), (f:$\Box$), (g:$\bigtriangledown$), (h:$\bullet$), (i:$\diamond$), (j:$\bigtriangleup$) and (k:$\times$).}  \label{fig:colecole}
\end{figure}

\begin{figure}[t]
\centering
  \includegraphics[width=6in]{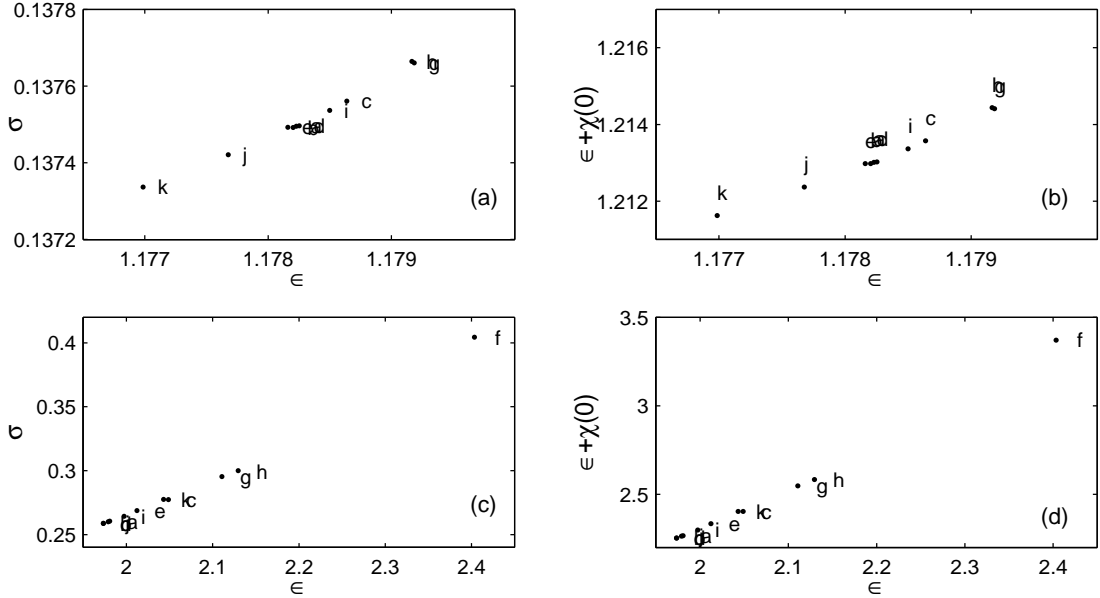}
  \caption{Frequency independent parameters for the match composites for considered concentration levels (a) and (b) $q=0.1$ and (c) and (d) $q=0.4$ for the structures in Fig.~\ref{fig:geometries}. (a) and (c) ohmic conductivity $\sigma$ versus permittivity at high frequencies $\epsilon$. (b) and (c) low frequency permittivity $\epsilon+\chi(0)$ versus permittivity at high frequencies $\epsilon$.}  \label{fig:epsilon}
\end{figure}
In the simulations the geometries in Fig.~\ref{fig:geometries}  are used, and the electrical properties of the structures are calculated with the finite element method \cite{Tuncer2002a,Tuncer2001a,TuncerAcc1}. The computational domains used in the simulations are marked in Fig.~\ref{fig:geometries} with rectangles. These regions are the smallest repeating units (unit-cells) of the structures. The phase parameters are chosen to correspond to match- and reciprocal composites as presented in Table~\ref{tab:matchreciprocal}. For descriptions of match and reciprocal composites see Refs.~\cite{Tuncer2002b,TuncerPhysD}. These parameters for the intrinsic electrical properties of phases yield relaxation times $\tau$ around $1\ \second$ for the interfacial polarization in a dilute mixture.
 
\begin{table}[b]
  \caption{Material parameters adopted in the finite element simulations.\label{tab:matchreciprocal}}  
  \begin{indented}
    \lineup
  \item[]  \begin{tabular}{lrr}
      \br
      Composite& $\varepsilon_1$ & $\varepsilon_2$ \\
      \mr
      match &$1-\imath\,(100\varepsilon_0\omega)^{-1}$ & $10-\imath\,(\varepsilon_0\omega)^{-1}$\\
      reciprocal &$1-\imath\,(\varepsilon_0\omega)^{-1}$ & $10-\imath\,(100\varepsilon_0\omega)^{-1}$\\
      \br
    \end{tabular}
  \end{indented}
\end{table}

\begin{figure}[t]
\centering
    \includegraphics[width=6in]{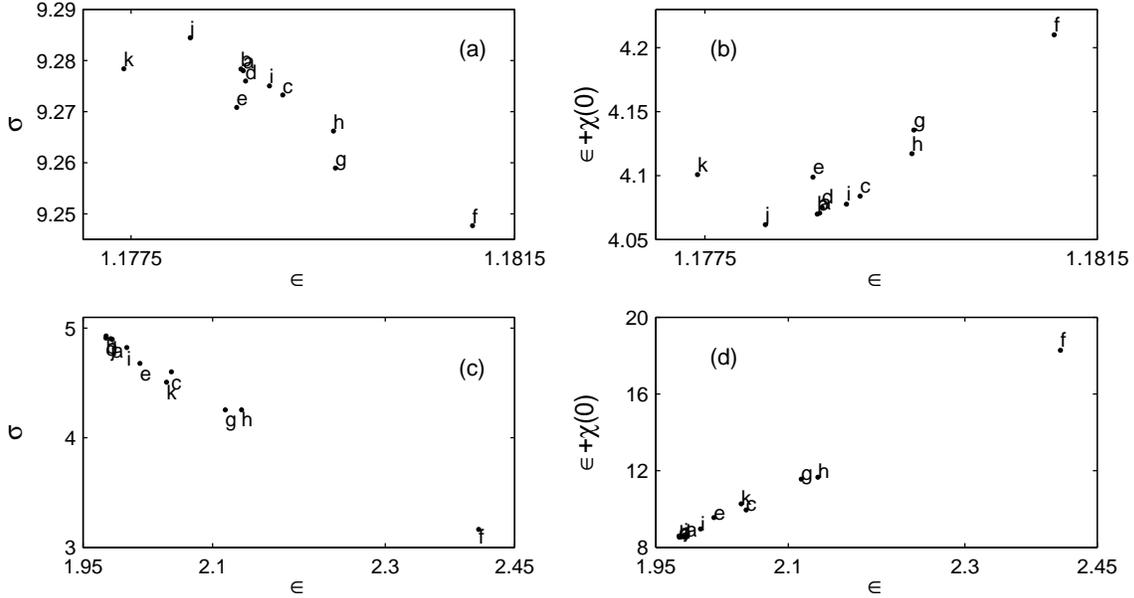}
\caption{Frequency independent parameters for the reciprocal composites for considered concentration levels (a) and (b) $q=0.1$ and (c) and (d) $q=0.4$ for the structures in Fig.~\ref{fig:geometries}. (a) and (c) ohmic conductivity $\sigma$ versus permittivity at high frequencies $\epsilon$. (b) and (c) low frequency permittivity $\epsilon+\chi(0)$ versus permittivity at high frequencies $\epsilon$.} \label{fig:epsilon_r}
\end{figure}

In Fig.~\ref{fig:colecole} dielectric permittivities calculated are plotted after the subtraction of the ohmic losses. At low concentrations $q=0.1$, it is not important if the composite is match (Fig.~\ref{fig:colecole}a) or reciprocal (Fig.~\ref{fig:colecole}c), dielectric dispersions of structures are almost the same except the structure with the lowest limiting concentration, Fig.~\ref{fig:geometries}f. Because of the low concentration, the polarization of each fiber is not affected by the neighboring fibers, so the resulting polarization of the whole structure/composite material is as if each fiber is alone, and does not influenced by the nearest neighbors polarization. However, for the structure in Fig.~\ref{fig:geometries}f, the influence of the neighboring fibers are pronounceable. At high concentrations of the fibers (phase 2), the differences between each structure start to be significant due to influence of the polarization of the neighbors in Fig.~\ref{fig:colecole}b and \ref{fig:colecole}d. It is not obvious to the naked-eye that all the calculated dispersions are deformed Debye\cite{Debye1945}. It has previously been shown that for two-dimensional composites~\cite{Tuncer2002b,Tuncer2001a} and in dispersive systems~\cite{NonDebye} non-Debye relaxations would occur. Therefore, the dispersion data are analyzed by the Havriliak-Negami\cite{HN} empirical formula. 
\begin{equation}
  \label{eq:hn}
   \overline{\varepsilon}_{\sf hn}(\omega)=\epsilon+{\chi(0)}[1+(\imath\omega\tau_1)^{\alpha}]^{-\beta}+ \sigma(\imath\varepsilon_o\omega)^{-1}
\end{equation}
Here, $\alpha$ and $\beta$ are fitting parameters that define the shape of the dispersion, $\tau$ and $\chi(0)$ are the position (relaxation time) and its amplitude (dielectric strength), respectively. This equation is the same as Eq.~(\ref{eq:1}) $\overline{\chi}(\omega)$ being the Havriliak-Negami empirical formula. In Tables~\ref{tab:matchHN} and \ref{tab:reciprocalHN}, Havriliak-Negami fits are presented together with the fitting errors. The error is calculated from the following expression,
\begin{equation}
  \label{chisquare}
  {\sf error}=\sum\{[\Re(\varepsilon_{\sf e}-\varepsilon_{\sf hn})/\Re(\varepsilon_{\sf e})^2]+[\Im(\varepsilon_{\sf e}-\varepsilon_{\sf hn})/\Im(\varepsilon_{\sf e})^2]\}
\end{equation}
The calculated errors indicate the fitness of the model function in Eq.~(\ref{eq:hn}). Only at high concentrations, the structures with low limiting concentrations $q_{\sf m}$ or in other words low percolation threshold show poor fitting results, which indicate that actually the Havriliak-Negami equation is not suitable for such an analysis. Although some of the $\beta$ values are over 1, the dispersion satisfies the condition $\alpha\le1$ and $\alpha\beta\le1$~\cite{TuncerPhysD}. The relaxation times $\tau$ obtained from the curve fittings for match composites at low concentrations are approximately the same. The shape of the relaxations are also Debye-like. However as the concentration is increased due to the structural differences and the interaction between inclusions, each geometry has a significant relaxation time $\tau$. Moreover, the shape parameters of the Havriliak-Negami empirical formula changes. As we consider the fitting results of the reciprocal composites, they illustrate significant differences between the structures. Keep in mind that the concentration of the inclusions are the same in Tables~\ref{tab:matchHN} and \ref{tab:reciprocalHN}, the span of the relaxation times $\tau$ and $\chi(0)$ show the influence of the structure. If we have the data as a black box, it would not be possible to calculate the concentration of the each phase as presented in the previous section, however, if we instead use the spectral density representation we would obtain valuable information regarding the structure of composite as presented in Table~\ref{tablepositionx}. The relaxation times of the match and reciprocal composites are altered and they are faster for the reciprocal composites due to more conductive matrix phase. Finally, the main difference between the reciprocal and match composites are the size of the dielectric strength which is higher for the former case because of the Maxwell-Wagner-Sillars polarization.

\begin{table}[t]
  \caption{Havriliak-Negami curve fitting parameters for the considered match composite structures. The last column is the error calculated using Eq.\ref{chisquare}.\label{tab:matchHN}}  
  \begin{indented}
    \lineup
  \item[]  \begin{tabular}{crrrrrrr}
      \br
      Fig.~\ref{fig:geometries}& $\epsilon$ & $\sigma$ &$\alpha$ &$\beta$ &$\tau$ &$\chi(0)$ & ${\sf error}$\\
&[$\varepsilon_0]$&$[\pico\siemens]$ &&&$[\second]$ & &\\
      \mr
      \centre{8}{ $q=0.1$}\\
      \mr
      a& 1.18 & 1.22& 0.994&  1.01& 0.983& 0.035& 7.7$\times10^{-7}$\\
      b& 1.18 & 1.22& 0.999&     1& 0.982& 0.0348& 2.1$\times10^{-8}$\\
      c& 1.18 & 1.22& 0.999&     1& 0.982& 0.035& 4.7$\times10^{-8}$\\
      d& 1.18 & 1.22& 0.992&  1.01& 0.985& 0.0351& 1.7$\times10^{-6}$\\
      e& 1.18 & 1.22& 0.996&     1& 0.984& 0.035& 3.6$\times10^{-7}$\\
      f& 1.18 & 1.22&     1&     1& 0.988& 0.0363& 1.4$\times10^{-8}$\\
      g& 1.18 & 1.22& 0.999&     1& 0.983& 0.0353& 2.1$\times10^{-7}$\\
      h& 1.18 & 1.22&     1&     1& 0.984& 0.0353& 1.8$\times10^{-8}$\\
      i& 1.18 & 1.22&     1& 0.999& 0.981& 0.0348& 3.7$\times10^{-8}$\\
      j& 1.18 & 1.22& 0.999&     1& 0.982& 0.0347& 2.5$\times10^{-8}$\\
      k& 1.18 & 1.22&     1&     1& 0.982& 0.0346& 2.9$\times10^{-8}$\\
      \mr
      \centre{8}{ $q=0.4$}\\
      \mr
      a& 1.98 & 2.31&     1&     1&  1.07& 0.286& 1.7$\times10^{-7}$\\
      b& 1.97 & 2.29&     1&     1&  1.07& 0.279& 1.3$\times10^{-7}$\\
      c& 2.05 & 2.46&     1& 0.997&  1.13& 0.354& 3.1$\times10^{-6}$\\
      d& 1.97 & 2.29&     1& 0.995&  1.07& 0.279& 3.9$\times10^{-6}$\\
      e& 2.01 & 2.38& 0.998&     1&  1.11& 0.322& 2.3$\times10^{-7}$\\
      f&  2.4 & 3.58& 0.989&  1.01&  1.55& 0.975& 5.2$\times10^{-5}$\\
      g& 2.11 & 2.62& 0.998&     1&  1.19& 0.437& 1.6$\times10^{-7}$\\
      h& 2.13 & 2.66& 0.998&     1&   1.2& 0.455& 1.8$\times10^{-7}$\\
      i&    2 & 2.34&     1&     1&  1.09& 0.301& 7.4$\times10^{-8}$\\
      j& 1.98 &  2.3&     1&     1&  1.07& 0.284& 9.4$\times10^{-8}$\\
      k& 2.04 & 2.46& 0.998&     1&  1.14& 0.361& 2.1$\times10^{-7}$\\
      \br
    \end{tabular}
  \end{indented}
\end{table}
\begin{table}[t]
  \caption{Havriliak-Negami curve fitting parameters for the considered reciprocal composite structures. The last column is the {\em chi-square} error calculated using Eq.\ref{chisquare}.\label{tab:reciprocalHN}}  
  \begin{indented}
    \lineup
  \item[]  \begin{tabular}{crrrrrrr}
      \br
      Fig.~\ref{fig:geometries}& $\epsilon$ & $\sigma$ &$\alpha$ &$\beta$ &$\tau$ &$\chi(0)$ & ${\sf error}$\\
&$[\varepsilon_0]$&$[\pico\siemens]$ &&&$[\second]$ & &\\
      \mr
      \centre{8}{ $q=0.1$}\\
      \mr
      a& 1.18 & 82.1&     1&     1& 0.806&  2.89& 1.0$\times10^{-6}$\\
      b& 1.18 & 82.1& 0.999&     1& 0.806&  2.89& 7.7$\times10^{-7}$\\
      c& 1.18 & 82.1& 0.998&     1& 0.807&  2.91& 5.5$\times10^{-7}$\\
      d& 1.18 & 82.1& 0.999&     1& 0.807&   2.9& 3.9$\times10^{-7}$\\
      e& 1.18 & 82.1& 0.998&     1& 0.814&  2.92& 3.3$\times10^{-6}$\\
      f& 1.18 & 81.9& 0.974&  1.03& 0.829&  3.04& 3.5$\times10^{-5}$\\
      g& 1.18 &   82& 0.991&  1.01& 0.817&  2.96& 3.6$\times10^{-6}$\\
      h& 1.18 &   82& 0.991&  1.01& 0.813&  2.94& 3.2$\times10^{-6}$\\
      i& 1.18 & 82.1& 0.999&     1& 0.807&   2.9& 1.4$\times10^{-7}$\\
      j& 1.18 & 82.2&     1&     1& 0.806&  2.88& 4.3$\times10^{-7}$\\
      k& 1.18 & 82.1& 0.994&  1.01& 0.817&  2.93& 2.1$\times10^{-6}$\\
      \mr
      \centre{8}{ $q=0.4$}\\
      \mr
      a& 1.98 & 43.4& 0.979&  1.03& 0.466&   6.7& 3.7$\times10^{-5}$\\
      b& 1.97 & 43.6& 0.999&     1& 0.466&  6.56& 3.5$\times10^{-7}$\\
      c& 2.05 & 40.8& 0.834&  1.24& 0.482&  8.09& 2.6$\times10^{-3}$\\
      d& 1.97 & 43.5& 0.992&  1.01& 0.468&  6.63& 5.1$\times10^{-6}$\\
      e& 2.01 & 41.4& 0.884&  1.16& 0.485&  7.64& 9.9$\times10^{-4}$\\
      f& 2.41 &   28&  0.76&  1.16&  1.78&  17.3& 7.2$\times10^{-2}$\\
      g& 2.11 & 37.7& 0.748&  1.39& 0.556&  9.95& 1.2$\times10^{-2}$\\
      h& 2.13 & 37.7& 0.734&  1.42& 0.551&  10.1& 1.4$\times10^{-2}$\\
      i&    2 & 42.7&  0.94&  1.08& 0.468&     7& 2.5$\times10^{-4}$\\
      j& 1.98 & 43.4& 0.983&  1.02& 0.467&  6.67& 2.1$\times10^{-5}$\\
      k& 2.04 & 39.9& 0.831&  1.24& 0.512&  8.44& 2.9$\times10^{-3}$\\
      \br
    \end{tabular}
  \end{indented}
\end{table}
In Fig.~\ref{fig:epsilon} and \ref{fig:epsilon_r} frequency independent parameters of the composites are presented to give some more information on the structures and to be used to identify the dispersions in Fig.~\ref{fig:colecole}. For match composites at low concentrations there are small dissimilarities, illustrated in Fig.~\ref{fig:epsilon}a and \ref{fig:epsilon}b. As presented previously when the concentration of the inclusions in the composite systems get close to the percolation threshold the structural differences dominate the electrical properties. This is shown  in Fig.~\ref{fig:epsilon}c and \ref{fig:epsilon}d, in which structures in Fig.~\ref{fig:geometries}f,~\ref{fig:geometries}g and  ~\ref{fig:geometries}h separate from the others. When the conductivities of the phases are interchanged and a reciprocal composite is formed even at low concentrations. The discrepancy between different structures are significant this is due to the increase of polarized charge at the interface. 


\begin{figure}[t]
\center{
\includegraphics[]{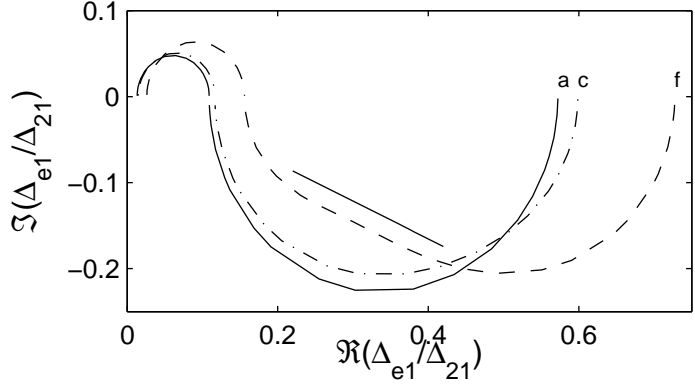}}
\caption{Parametric representation of the scaled composite permittivity for selected three structure from Fig.~\ref{fig:geometries}. The negative imaginary parts are obtained for reciprocal composites. \label{fig:delta}}
\end{figure}

Once the dielectric dispersions are known together with the intrinsic dielectric properties of the constituents, the data can be converted to the spectral density representation form, the scaled dielectric permittivities $\Delta_{ij}$ in Eq.~(\ref{eq:delta}). This is illustrated in Fig.~\ref{fig:delta}. It is graphical to see the structural differences in the scaled permittivities. The smaller semi-circles in the positive quadrant are for match composites for $q=0.4$. The negative imaginary parts are obtained for reciprocal composites for the same concentration. It is significant that Fig.~\ref{fig:geometries}a and \ref{fig:geometries}c change behavior as the real part of the scaled permittivity increases. Moreover, it is similar to Fig.~\ref{fig:colecole}, that the structure with the lowest $q_{\sf m}$ has the largest deviation of all, one very significant feature of this structure is that the flat part marked with the solid line (\full) in the figure, which illustrates the deviation from the simple Debye relaxation. As a note, the Debye relaxation observed in layered-structures correspond to a pole at $x=1$ or $x=0$ in the spectral representation, which is actually a delta function $\delta(x-1)$ or $\delta(x)$ depending on the direction of the applied field and the layered structure.
  
\begin{figure}[t]
\centering
  \begin{tabular}{lr}
  \includegraphics[]{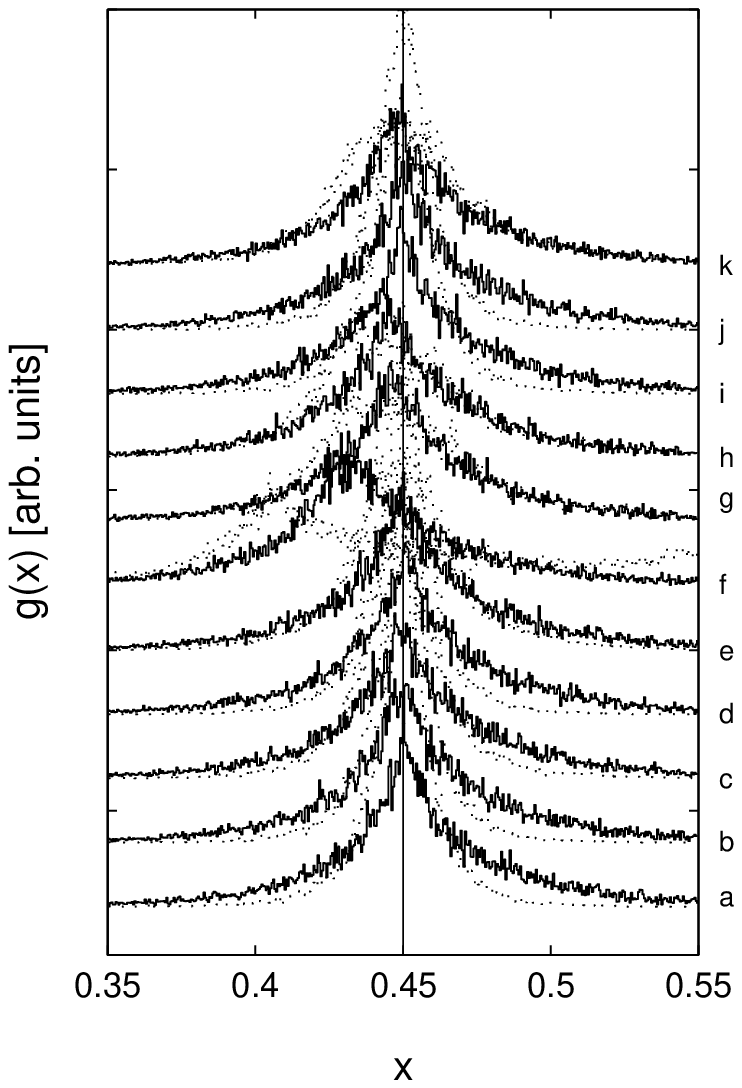}&
  \includegraphics[]{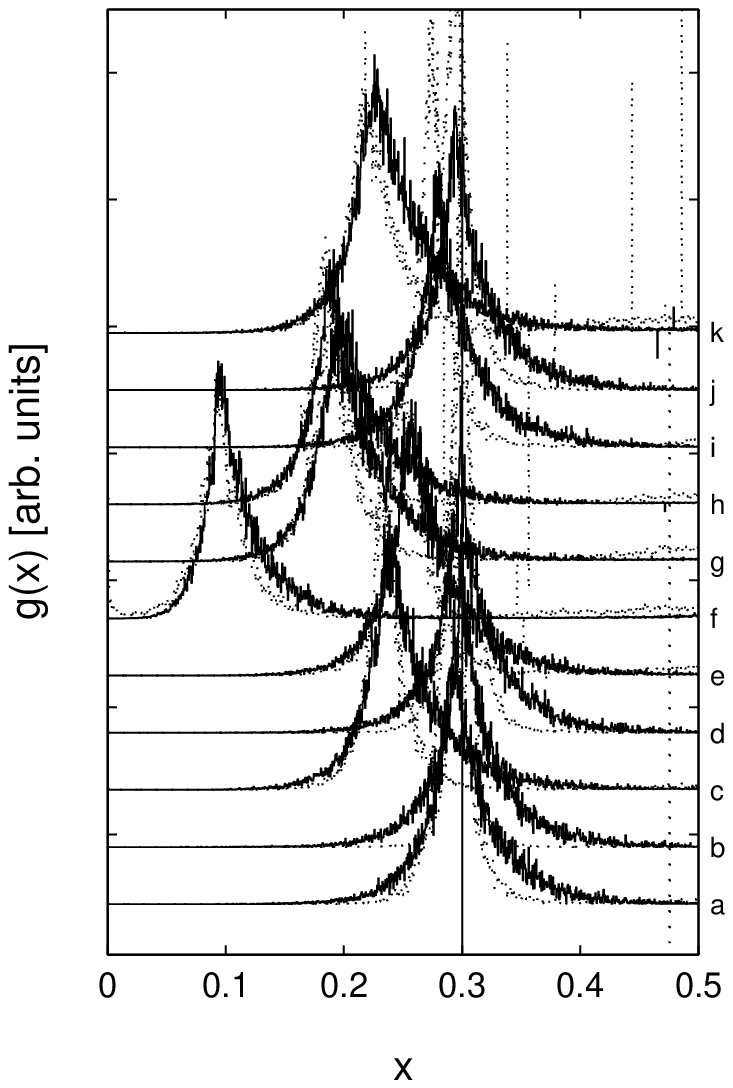}
  \end{tabular}
\caption{Calculated spectral density distributions. (a) At low concentrations density distributions are similar and like delta sequences. (b) At high concentrations, distributions are broader. The solid lines are the expected positions for the spectral density distributions from the effective medium theory Eq.~(\ref{eq:3}). The solid and dotted lines are calculations from match and reciprocal composites, respectively.}  \label{fig:spectrum}
\end{figure}

The analysis of the scaled permittivities with the help of the developed numerical method resulted  in the spectral density functions presented in Fig.~\ref{fig:spectrum}. It is clear that the structure in Fig.~\ref{fig:geometries}f is different than the others at low concentrations, Fig.~\ref{fig:spectrum}a, because its peak is not located near the solid line (the expected position for the depolarization factor from the Maxwell-Garnett expression). Although it is previously stated many times in the present text,  all the other structures yield the same spectral distributions, emphasizing that they are not actually different. This is also a verification of the effective medium approach. (A similar result can also be drawn from Fig.~\ref{fig:colecole}a and \ref{fig:colecole}c.) When the high concentration case ($q=0.4$) is taken into consideration, Fig.~\ref{fig:spectrum}b, there are significant differences between the structures. Now only geometries Fig.~\ref{fig:geometries}b, \ref{fig:geometries}d and \ref{fig:geometries}j yield similar spectral density distributions, which coincide with the suggested depolarization factor of the effective medium theory [solid line (\full) and Eq.~(\ref{eq:3})]. This reveals that these structures are in fact not close to the percolation threshold. The other structures with lower percolation thresholds ($q_{\sf m}<0.8$) spread in the spectrum. The structure in Fig.~\ref{fig:geometries}f approaches to zero depolarization factor, giving a sign for start of a percolation--touching of fibers to form a continues path from one side to other. In Table~\ref{tablepositionx}, calculated positions of the structural parameters are presented. There are little discrepancies between values obtained whether the composite is match or reciprocal. The table also verifies the statements regarding position of the depolarization factors, $\overline{x}$ and the percolation thresholds. 

\begin{table}[t]
\caption{Peak positions of depolarization factors $\overline{x}$ calculated with the L{\'e}vy statistics for the geometries in Fig.~\ref{fig:geometries}. The numerical  method to extract the spectral density function is applied both to match and reciprocal composite configurations. The expected values for depolarization factor from the effective medium theory for $q=0.1$ and $q=0.4$ are  $0.45$ and $0.3$ [calculated from Eq.~(\ref{eq:3}) with $d=2$], respectively.} \label{tablepositionx}
\begin{indented}
\lineup
\item[]\begin{tabular}{c@{~~~~}*{4}{c}}
\br
&\centre{2}{$q=0.1$} & \centre{2}{$q=0.4$}\\
Fig.~\ref{fig:geometries} & match & reciprocal & match & reciprocal \\
\mr
a&0.4513 & 0.4477 & 0.2961 & 0.2892 \\
b&0.4511 & 0.4477 & 0.3013 & 0.2964 \\
c&0.4503 & 0.4481 & 0.2443 & 0.2323 \\
d&0.4512 & 0.4481 & 0.3010 & 0.2963 \\
e&0.4505 & 0.4493 & 0.2642 & 0.2540 \\
f&0.4348 & 0.4119 & 0.0980 & 0.0900 \\
g&0.4473 & 0.4466 & 0.2008 & 0.1886 \\
h&0.4466 & 0.4460 & 0.1958 & 0.1853 \\
i&0.4504 & 0.4480 & 0.2840 & 0.2731 \\
j&0.4508 & 0.4478 & 0.2965 & 0.2921 \\
k&0.4493 & 0.4497 & 0.2358 & 0.2219 \\
\br
\end{tabular}
\end{indented}
\end{table}

\section{Discussions}

In this section, I would like to take three examples and with the help of the information presented in the previous section characterize the differences between structures. For this analysis as described in \S~\ref{sec:import-mixt-topol}, the structures in Figs.~\ref{fig:geometries}a and \ref{fig:geometries}d, Figs.\ref{fig:geometries}g and \ref{fig:geometries}h, and Figs.\ref{fig:geometries}c, \ref{fig:geometries}i and \ref{fig:geometries}k are selected. 

First, let us consider the structures in Figs.~\ref{fig:geometries}a and \ref{fig:geometries}d, which have nearly the same percolation threshold but the number of the nearest neighbors, Table~\ref{tab:table1}. Since the considered concentration values are away from the percolation thresholds, the Havriliak-Negami empirical formula and the spectral representation approach resulted in a very similar way for the both concentration cases for the match composites and low concentrations of reciprocal composite arrangements. For high concentrations $q=0.4$, the reciprocal composites of these structures have slightly different  curve fitting parameters as presented in Table~\ref{tab:reciprocalHN}, and the errors in the fitting are higher if compared to the ones at low concentrations, Tables~\ref{tab:matchHN} and \ref{tab:reciprocalHN}.  The depolarization factors from the spectral density analyses in Fig.~\ref{fig:spectrum} and Table~\ref{tablepositionx} on the other hand illustrates minor differences for both match and reciprocal cases. The spectral representation states that the obtained spectrum contains contribution of the pure geometry, it is clear that we are not able to see the difference between these two structure when we compare the obtained depolarization functions, Table~\ref{tablepositionx}.

Secondly, we concentrate on the structures in Figs.~\ref{fig:geometries}g and \ref{fig:geometries}h, which have similar percolation threshold and the same number of the nearest neighbors.  For both simulation considerations, match and reciprocal, at low concentrations these structures yield similar dielectric dispersions, see Tables~\ref{tab:matchHN} and \ref{tab:reciprocalHN}. However, at high concentration of inclusions $q=0.4$, only the relaxation shape parameters $\alpha$ and $\beta$ and the position $\tau$ of the dispersions are alike in the curve curve fitting analysis with Eq.~(\ref{eq:hn}). The dispersions amplitude $\chi(0)$, permittivity at high frequencies $\epsilon$ and ohmic conductivity $\sigma$ differ from each other. For reciprocal composite case the calculated error is large, however, the same is not true for the match composite, therefore the Havriliak-Negami empirical formula can be used to differentiate differences between structures. However, the spectral representation approach as presented in Table~\ref{tablepositionx} yield slightly different values for the depolarization factors, which indicate that the structures are separable. We can also speculate that the differences in the frequency independent parameters might have been due to the influence of the second nearest neighbors. One other explanation can be the long range interaction of the clusters of inclusions, this can be seen as the change in the orientation of inclusions in groups of four in Figs.~\ref{fig:geometries}g and \ref{fig:geometries}h. 

Finally, the three structures in Figs.~\ref{fig:geometries}c and  \ref{fig:geometries}i and \ref{fig:geometries}k have the same percolation threshold ($q=0.61$) but different coordination numbers, 3, 5 and 4 respectively (Table~\ref{tab:table1}). At low concentrations far away from $q_{\sf m}$, these composites yield the same spectral density function. The application of the Havriliak-Negami function also confirms the findings. At high concentration, $q=0.4$, the differences between the structure become visible in both analyses, the spectral density representation and the Havriliak-Negami curve-fitting. The match and reciprocal composite cases indicate that the structure with the largest coordination number (Fig.~\ref{fig:geometries}i) yield low permittivity values compared to the others, Tables~\ref{tab:matchHN} and \ref{tab:reciprocalHN}. When the conductivity is considered, the same is observed for the match composite, Table~\ref{tab:matchHN}, which is expected due to the similarity of the capacitive and resistive problem. The reciprocal composite of the same structure on the contrary results the opposite and yield a highest conductivity $\sigma$ of all three, Table~\ref{tab:reciprocalHN}. The two structures Figs.~\ref{fig:geometries}c and ~\ref{fig:geometries}k have similar electrical properties, when the error in the curve-fitting are taken into consideration in Tables~\ref{tab:matchHN} and \ref{tab:reciprocalHN}, the differences are within the error. The curve fittings reveal that the dispersions obtained are non-Debye type and depending on the intrinsic properties of the phases can be symmetrical or non-symmetrical.

\begin{table}[t]
\caption{A summary of the dielectric relaxation in binary composite depending on the intrinsic electrical properties of constituents and topology. The inclusions in the simulations on regular structures have been mono-dispersed as in the present study. The deduction are from the simulations performed by the author. The concentration low and high means relation to the limiting concentration $q_{\sf m}$ in regular topologies and percolation threshold in the disordered ones. S-reciprocal is the slightly reciprocal case, and E-reciprocal is the extreme reciprocal case. For E-reciprocal composites, dielectric dispersion of the system exhibits in low-frequency-dispersion (lfd) or in other words constant phase angle behavior~\cite{TuncerPhysD}.}  \label{tab:general}
\begin{indented}
\lineup
\item[]\begin{tabular}{cccc}
    \br
    concentration&case&topology&relaxation\\
    \mr
    low  & match    & regular & Debye\\
    low  & reciprocal  & regular & Debye\\
    low  & match    & random & Debye\\
    low  & reciprocal    & random & Debye\\
    high  & match & regular & symmetric\\
    high  & S-reciprocal  & regular & non-symmetric\\
    high  & S-reciprocal & random & non-symmetric\\
    high  & E-reciprocal  & random & lfd\\
    \br
  \end{tabular}
\end{indented}
\end{table}

As a remark, the dielectric dispersion in two-dimensional binary composites can be summarized as in Table~\ref{tab:general}. The items in the table have been obtained purely from numerical simulations,  published previously elsewhere~\cite{TuncerPhysD,TuncerRandom,Tuncer2002a,Tuncer2002b,Tuncer2001a,Tuncer2002elec}. At low concentrations of inclusions the dielectric dispersion are in the Debye form, and they can be analyzed by the effective medium theories for a given shape of inclusion, when the inclusions are mono-dispersed. The term ``low concentration'' in Table~\ref{tab:general} is meant that the concentration of inclusions is relative to the limiting concentration, the level at which the inclusions start to  touch each other and form a continues chain from one-side of the composite to the other one. In the disordered/random structures, the low concentration cases would then be assigned relative to the percolation threshold. In disordered/random composites the dielectric dispersions are non-Debye type, and depending on the intrinsic electrical properties of the constituents even low frequency dispersion (lfd) or also known as constant phase angle~\cite{MacDonald1987,Jonscher1983,jons92,Ngai1979,Dissado1,IshidaHill,HillLFD} type of dispersions are obtained on a wide-frequency range. The frequency range of the lfd dispersions can be increased if the number of components (inclusion phase) or the largest to smallest length scale ratio  are increased in the numerical simulations~\cite{TuncerPhysD}. Although, only the position of the depolarization factors are presented and considered in the data analysis and discussions, the shape of the spectral functions also provide information regarding the micro-structure of the composite. As stated by Lysne~\cite{Lysne1983}, the position of the depolarization factor in the spectral density representation is correlated with the relaxation time of the dielectric dispersion. Therefore, one can in principle use the distribution of relaxation times~\cite{Tuncer2004a,Tuncer2000b}, however, it would be cumbersome, since there is no trivial quantity as the depolarization factor in the spectral density representation to correlate the relaxation times. 

\section{Conclusions}

In this paper, an insight to the electrical properties of binary composites are given with the help of numerical simulations. It is presented that if we are able to control the manufacturing process in preparing materials for desired electrical properties, the computer simulation help us not only to characterize but to design materials as well. Eleven selected space filling geometries (lattices) are used in the numerical simulations. The structures are ``ideal composites'', in which we can change the intrinsic properties and distribution of inclusions. The dielectric response for each structure is calculated for two concentration levels of inclusions. The obtained dielectric dispersions are analyzed by the Havriliak-Negami empirical formula and the spectral density representation method. It is concluded that for low concentrations of inclusions (lower than the limiting concentration or the percolation threshold for a given system), the composite systems exhibit Debye-type dispersions, which can be modeled by empirical formulas for mixtures. At concentrations close to the percolation threshold, the structures illustrate deviations from the Debye-type relaxation. The spectral density approach illustrate some advantages over the effective medium approach and using empirical formulas for the dielectric dispersions obtained when the systems have high concentrations of inclusions. It is now time to create look-up tables for high concentration composites with known micro-structures in order to compare  experimental and computer simulated dielectric dispersion of composites. With the help of the look-up tables, we would be able to know the internal structure, such as the distribution of inclusions. Thus, the dielectric spectroscopy data on composites together with the spectral density representation can be used as a probe to provide quantitative information on the micro-structure of the composite.

As a scheme or a next step, it is valuable to prepare composites with known internal structures and verify experimental results with numerical simulations or the opposite. With the improving material preparation techniques, this should be a task for both theorists and experimentalists. 
\bibliography{onmicro.bib}
\bibliographystyle{unsrtnat}
\end{document}